\documentclass[10pt,conference]{IEEEtran} 
\IEEEoverridecommandlockouts
\usepackage{cite}
\usepackage{amsmath,amssymb,amsfonts}
\usepackage{algorithmic}
\usepackage{graphicx}
\usepackage{textcomp}
\usepackage{xcolor}
\usepackage{booktabs}

\def\BibTeX{{\rm B\kern-.05em{\sc i\kern-.025em b}\kern-.08em
    T\kern-.1667em\lower.7ex\hbox{E}\kern-.125emX}}

    \usepackage{scalerel}
\usepackage{tikz}
\usetikzlibrary{svg.path}

\definecolor{orcidlogocol}{HTML}{A6CE39}
\tikzset{
  orcidlogo/.pic={
    \fill[orcidlogocol] svg{M256,128c0,70.7-57.3,128-128,128C57.3,256,0,198.7,0,128C0,57.3,57.3,0,128,0C198.7,0,256,57.3,256,128z};
    \fill[white] svg{M86.3,186.2H70.9V79.1h15.4v48.4V186.2z}
                 svg{M108.9,79.1h41.6c39.6,0,57,28.3,57,53.6c0,27.5-21.5,53.6-56.8,53.6h-41.8V79.1z M124.3,172.4h24.5c34.9,0,42.9-26.5,42.9-39.7c0-21.5-13.7-39.7-43.7-39.7h-23.7V172.4z}
                 svg{M88.7,56.8c0,5.5-4.5,10.1-10.1,10.1c-5.6,0-10.1-4.6-10.1-10.1c0-5.6,4.5-10.1,10.1-10.1C84.2,46.7,88.7,51.3,88.7,56.8z};
  }
}

\newcommand\orcidlink[1]{\href{https://orcid.org/#1}{\mbox{\scalerel*{
\begin{tikzpicture}[yscale=-1,transform shape]
\pic{orcidlogo};
\end{tikzpicture}
}{|}}}}

\usepackage[hidelinks]{hyperref}

\begin{document}

\title{Vulnerability Propagation in Package Managers Used in iOS Development
\thanks{
Funding of this research came from grant PRG1226 of the Estonian Research Council, the European Social Fund via the IT Academy program, 
the Federal Ministry for Climate Action, Environment, Energy, Mobility, Innovation and Technology (BMK), the Federal Ministry for Digital and Economic Affairs (BMDW), and the State of Upper Austria in the frame of the SCCH competence center INTEGRATE (FFG grant no. 892418) part of the COMET - Competence Centers for Excellent Technologies Programme managed by Austrian Research Promotion Agency FFG.
}
}

\author{\IEEEauthorblockN{1\textsuperscript{st} Kristiina Rahkema \orcidlink{0000-0001-7332-2041}}
\IEEEauthorblockA{\textit{Institute of Computer Science} \\
\textit{University of Tartu}\\
Tartu, Estonia \\
kristiina.rahkema@ut.ee}
\and
\IEEEauthorblockN{2\textsuperscript{nd} Dietmar Pfahl \orcidlink{0000-0003-2400-501X}}
\IEEEauthorblockA{\textit{Institute of Computer Science} \\
\textit{University of Tartu}\\
Tartu, Estonia \\
dietmar.pfahl@ut.ee}
}

\maketitle

\begin{abstract}
Although using third-party libraries is common practice when writing software, vulnerabilities may be found even in well-known libraries. Detected vulnerabilities are often fixed quickly in the library code. The easiest way to include these fixes in a dependent software application, is to update the used library version. Package managers provide automated solutions for updating library dependencies, which make this process relatively easy. However, library dependencies can have dependencies to other libraries resulting in a dependency network with several levels of indirections. Assessing vulnerability risks induced by dependency networks is a non-trivial task for software developers.

The library dependency network in the Swift ecosystem encompasses libraries from CocoaPods, Carthage and Swift Package Manager. These three package managers are used while developing, for example, iOS or Mac OS applications in Swift or Objective-C. We analysed how vulnerabilities propagate in the library dependency network of the Swift ecosystem, how vulnerable dependencies could be fixed via dependency upgrades, and if third party vulnerability analysis could be made more precise given public information on these vulnerabilities.

We found that only 5.9\% of connected libraries had a direct or transitive dependency to a vulnerable library. Although we found that most libraries with publicly reported vulnerabilities are written in C, the highest impact of publicly reported vulnerabilities originated from libraries written in native iOS languages, i.e., Objective-C and Swift. We found that around 30\% of vulnerable dependencies could have been fixed via upgrading the library dependency. In case of critical vulnerabilities and latest library versions, over 70\% of vulnerable dependencies would have been fixed via a dependency upgrade. Lastly, we checked whether the analysis of vulnerable dependency use could be refined using publicly available information on the code location (method or class) of a reported vulnerability. We found that such information is not available most of the time.


\end{abstract}

\begin{IEEEkeywords}
iOS, Swift, Vulnerabilities, Library Dependency Networks, Publicly Reported Vulnerabilities
\end{IEEEkeywords}

\section{Introduction}\label{sec:introduction}

Using third-party libraries is common practice in software development. Third-party libraries make it possible to reuse existing solutions to common problems. This can make the development process faster and easier. These third-party solutions are often better vetted than custom solutions. The Open Web Application Security Project (OWASP), for example, strongly recommends against the use of custom encryption algorithms\cite{owasp-crypto}.

Nevertheless, vulnerabilities can be found in even very popular and well-tested libraries. For example, in December 2021, a security vulnerability was discovered in the widely used Log4J Java logging library. This vulnerability affected 4\% of all the Java applications \cite{google-neo4j} and made them vulnerable to remote code execution attacks. 

Many of these vulnerabilities are fixed relatively quickly \cite{zerouali2022impact}. After a fix is made available, dependents of the vulnerable library can include the fix via upgrading the library dependency version. It can, however, be tedious to update multiple dependencies manually. Automated solutions make this process easier. For this purpose, package managers have been created where the developer simply states the library name and exact version or a version requirement. The package manager takes care of downloading and installing the suitable library version. 

Using a package manager, it is easy to declare as many dependencies as needed. These library dependencies themselves can, again, have dependencies to other libraries, creating a network of library dependencies. The collection of all libraries that are available through a package manager and their library dependencies create a library dependency network for each package manager.
When the number of direct and transitive dependencies (i.e. indirect dependencies of any level of indirection) grows, it also increases the risks of a library depending on vulnerable library versions. As seen in the example of Log4J, even a vulnerability in a seemingly harmless logging library can have an affect on a significant part of an ecosystem. 

The spread of vulnerabilities in package manager library dependency networks has been studied for some package managers. Zerouali et al. \cite{zerouali2022impact} studied how long it takes for vulnerabilities in npm and RubyGems to be fixed and how these vulnerabilities spread through the library dependency network. They found that around 40\% of libraries have a direct or transitive dependency to a vulnerable library version.  Düsing et al. \cite{dusing2021analyzing} analyzed how vulnerabilities in transitive dependencies affect the NuGet, npm and Maven package manager library dependency networks. They also studied how fast developers update their library dependencies when a vulnerability is publicly disclosed. They found, that there is a significant difference on how many libraries are affected by vulnerable dependencies depending on the package manager. They also found, that developers probably rely on automated dependency updates, which are triggered when a vulnerability is disclosed. 

Although there are many studies that analyze library dependency networks, especially for npm and Maven, there are no studies analyzing the library dependency networks of CocoaPods, Carthage and Swift Package Manager (Swift PM). These three package managers are used when developing applications in Swift, such as iOS, Mac OS or Watch OS applications. In the following, we refer to the combined ecosystems of CocoaPods, Carthage and Swift PM as the Swift ecosystem. It is important to note that this ecosystem also contains libraries written in other languages (such as Objective-C, C, C++). Additionally, CocoaPods and Carthage can also be used in applications written in Objective-C. In the following, when referring to library dependency networks we mean the library dependency networks of the Swift ecosystem unless specified differently. 

Rahkema et al. \cite{rahkema2022checker} published a tool for finding publicly reported vulnerabilities in Swift application dependencies. The evaluation of the tool showed that there are indeed open source apps written in Swift that depend on library versions with publicly reported vulnerabilities. To better understand the magnitude of the use of vulnerable library versions, it would be necessary to analyse the spread of vulnerabilities in the entire library dependency network. No such studies have been conducted so far for the Swift ecosystem. In the following, we analyze how publicly reported vulnerabilities spread through the library dependency networks of the Swift ecosystem.

In our study, we investigate three research questions, focusing on (1) how the Swift ecosystem is affected by publicly reported vulnerabilities, (2) how risks from these vulnerabilities could be mitigated via dependency upgrades, and (3) if publicly available vulnerability information would allow more precise analysis of vulnerable dependencies.

The rest of the article is structured as follows. In Section \ref{sec:related}, we summarize related work. In Section \ref{sec:background}, we explain some background on the studied package managers, vulnerabilities and the dataset used. 
In Section \ref{sec:method}, we describe the research questions and how we analyzed the dataset for each research question. In Section \ref{sec:results}, we present answers to the tree research questions and Section \ref{sec:discussion} discusses these answers. In Section \ref{sec:threats}, we describe the threats to validity. In Section \ref{sec:conclusion}, we conclude the paper. 

\section{Related Work}\label{sec:related}

\subsection{Vulnerabilities in Library Dependency Networks}

Zerouali et al. \cite{zerouali2022impact} studied how long it takes for vulnerabilities in libraries from npm and RubyGems to be fixed, how vulnerabilities spread through the library dependency network and if vulnerable libraries are updated. 
They matched vulnerability data from Snyk to npm and RubyGems libraries and found that more than 15\% of latest library versions are directly dependent on vulnerable libraries. Additionally, dependencies to vulnerable libraries affected 42.1\% of npm and 39\% of RubyGems libraries. They found that one third of vulnerable dependencies could be fixed by updating the vulnerable dependency version.

Düsing et al. \cite{dusing2021analyzing} matched vulnerabilities from Snyk to libraries from the Maven, NuGet, and npm library dependency networks. They, then, analyzed how vulnerabilities in direct and transitive dependencies affect different library dependency networks. They found that only 1\% of libraries in NuGet and 8\% of libraries in npm are affected by vulnerable dependencies. Whereas, 29\% of libraries served through Maven have dependencies to vulnerable library versions. They also studied how long it takes for libraries to update their vulnerable dependencies after vulnerability disclosure and found, that at least some libraries are probably using automated tools that follow vulnerability databases and update all vulnerable dependencies automatically. 

Li et al. \cite{li2021pdgraph} analyzed library dependency networks of Java projects from Maven and GitHub. They matched vulnerability data from the National Vulnerability Database (NVD) to these Java projects and found 503 vulnerabilities matching 174 Maven projects and 3326 vulnerabilities matching 840 GitHub projects. They observed libraries with vulnerable dependencies from 2019 to 2020 and found that only 5\% of vulnerable dependencies were fixed during this time frame. Prana et al. \cite{prana2021out} analysed vulnerabilities in library dependencies for Java, Python and Ruby projects. They found that most vulnerabilities persisted through their one year long observation period.

Zimmermann et al. \cite{zimmermann2019small} studied security risks in the npm library dependency network. They found, that when installing an average npm library the user implicitly trusts 80 dependent libraries. When analyzing publicly reported vulnerabilities from Snyk, they found, that up to 40\% of all libraries have (direct or transitive) vulnerable dependencies. Alfadel et al. \cite{alfadel2020threat} analyzed the use of vulnerable npm dependencies in Node.js applications. They found, that although 67.9\% of examined applications depended directly on vulnerable libraries, 94.9\% of these vulnerabilities were not known at the time. 

Our study is the first study to analyse vulnerability propagation in the Swift ecosystem. 

\subsection{Vulnerability Reachability Analysis}
Tools have been implemented for multiple languages that allow more detailed analysis of vulnerable dependencies. Ponta et al \cite{ponta2020detection} implemented Eclipse Steady, that analyses if a vulnerable code in a dependency is called. The analysis relies on fix commits, which provided more accurate results than tools that relied only on vulnerability metadata such as OWASP Dependency Check. 

Bhandari et al. \cite{bhandari2021cvefixes} built a dataset containing vulnerable files and methods. Their analysis relied on the fixing commit being available on NVD. 

Hommersom et al. \cite{hommersom2021mapping} analysed how vulnerability fix commits can be found given public vulnerability information. They built a commit ranking system based on metrics such as time distance, commit messages and vulnerability description and others. They showed that detailed public vulnerability information can be successfully used to determine fixing commits, especially if the vulnerability description includes affected files, the patching commit or the commit message of the fixing commit refers to the CVE. 

In our study we analyse how often detailed information on the vulnerability location can be extracted from NVD, which would allow for a more detailed analysis of the vulnerable dependency. 

\section{Background}\label{sec:background}

In this section we describe the three package managers used in iOS development, how dependencies are declared, the available vulnerability information and the dataset used. 

\subsection{Package Managers}

There are three package managers that are used in iOS developement: CocoaPods, Carthage and Swift Package Manager (Swift PM). 

\textbf{CocoaPods} is the first package manager that was released in 2011\footnote{\url{https://cocoapods.org}}. CocoaPods has a centralized repository of libraries that anyone can add their libraries to. CocoaPods is easy to use, but somewhat heavyweight, as it forces its users to use a project file generated by the package manager. 

\textbf{Carthage} is the second oldest package manager in the Swift ecosystem and was released in 2014\footnote{\url{https://github.com/Carthage/Carthage}}. Carthage has no centralized list of libraries and is very lightweight. Developers using Carthage use the package manager to fetch and compile libraries, but the libraries need to be added to the projects manually. 

\textbf{Swift Packge Manager} is the latest and official package manager for the Swift language released in 2017\footnote{\url{https://www.swift.org/package-manager/}}. Swift PM also works as a project configuration file. Similarly to Carthage, Swift PM does not have a centralized list of libraries. When Swift PM was first released it only worked with command line applications. Since 2019 support for XCode and iOS has been added, making it the go to package manager in iOS development. 

\subsection{Dependency Declarations}

For each of the package managers dependencies are declared in a package manager manifest file. Once the package manager version resolution is run a resolution file is created that stores the actual library versions installed. The three package managers use different ways of declaring library dependencies, but work similarly in principle \cite{rahkema2022analysing}.

For example, for CocoaPods dependencies are declared in a Podfile. If Library C wanted to declare a dependency to Library B the Podfile would look as follows: 

\begin{verbatim}
    pod 'LibraryB'
\end{verbatim}

If Library B itself had a dependency to Library A, the resolution file Podfile.lock would include the following after CocoaPods installed the declared dependency: 

\begin{verbatim}
    PODS:
        - LibraryB (version1):
            - LibraryA 
        - LibraryA (version1)
\end{verbatim}

This package manager resolution file indicates that version 1 of Library B was installed and since Library B depends on Library A, version 1 of Library A was also installed. 

\begin{figure}[htbp]
\centerline{\includegraphics[width=1.0\columnwidth]{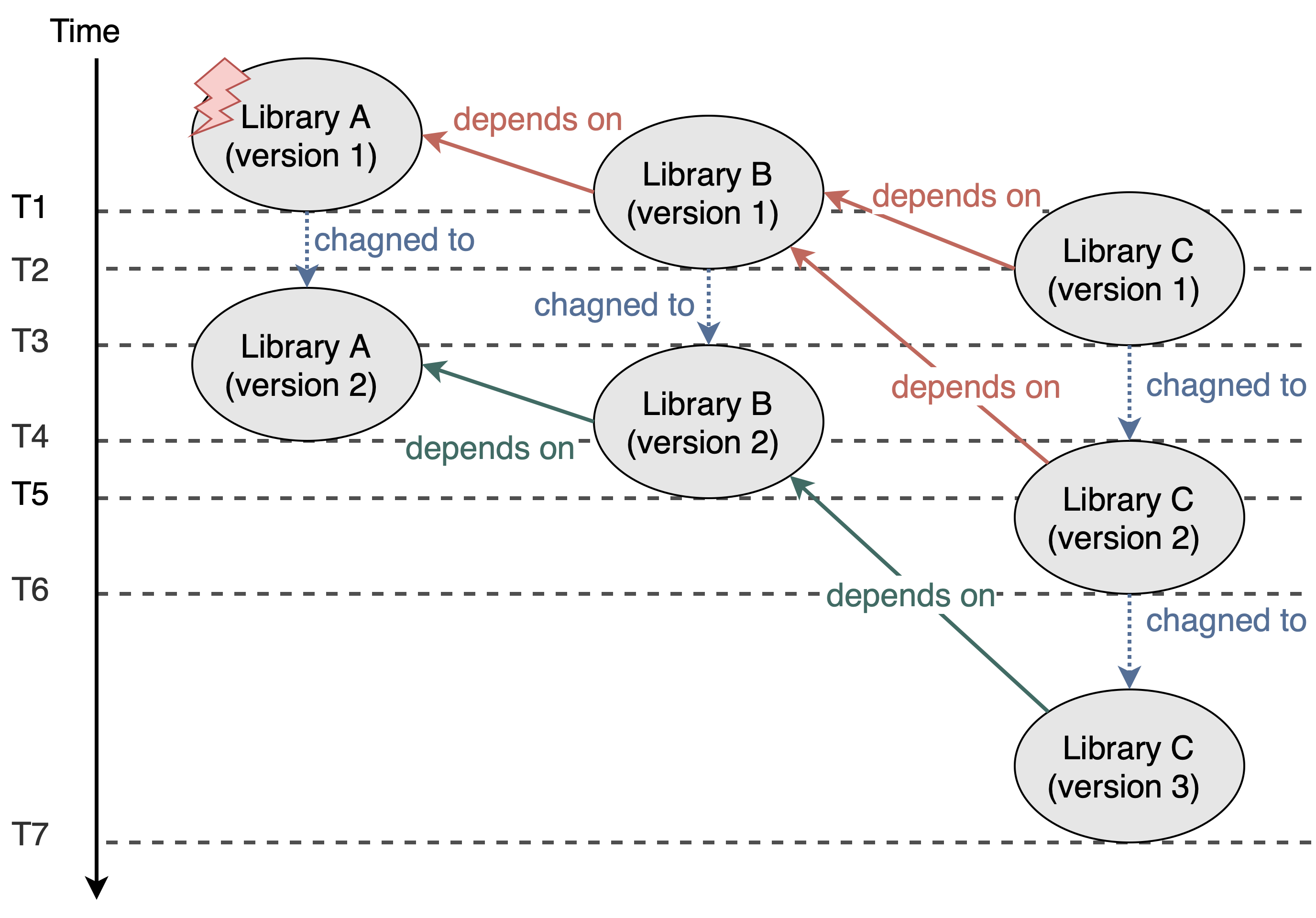}}
\caption{Illustration of dependency chains in a library dependency network with three libraries A, B and C.}
\label{fig:dependency-chains}
\end{figure}

There are now two dependency chains at time T3, as illustrated in Figure \ref{fig:dependency-chains}:
\begin{itemize}
    \item ABC1: A:v1 $\leftarrow$ B:v1 $\leftarrow$ C:v1
    \item AB1: A:v1 $\leftarrow$ B:v1
\end{itemize}

In Figure \ref{fig:dependency-chains}, version 1 of Library A has a publicly reported vulnerability. This means that dependency chains ABC1 and AB1 both represent dependencies to vulnerable library versions. At time T4, Library A releases a fix for the vulnerability: version 2 of Library A. At time T5, Library B releases a new version 2 that upgrades its library dependency to the not vulnerable version of Library A. At time T6, Library C releases a new version but does not upgrade its library dependency. At time T7, Library C releases another version that does update its library dependency to Library B version 2. At time T7, the following dependency chains exist in our dataset: 

\begin{itemize}
    \item ABC1: A:v1 $\leftarrow$ B:v1 $\leftarrow$ C:v1
    \item AB1: A:v1 $\leftarrow$ B:v1
    \item AB2: A:v2 $\leftarrow$ B:v2
    \item ABC2: A:v1 $\leftarrow$ B:v1 $\leftarrow$ C:v2
    \item ABC3: A:v2 $\leftarrow$ B:v2 $\leftarrow$ C:v3
\end{itemize}

Dependency chains ABC1, AB1 and ABC2 represent vulnerable dependencies. 

\subsection{Vulnerabilities}

If a vulnerability is discovered in a software system, it can either be fixed silently or the vulnerability can be released in public vulnerability databases. Such databases make it possible for users to monitor software systems they are using. One of such vulnerability databases is the National Vulnerability Database (NVD)\footnote{\url{http://nvd.nist.gov}}. 

NVD provides public information on vulnerabilities, such as a vulnerability description, severity levels, affected software, and others. Useful links are provided for some vulnerabilities, for example, to the security advisory or to the fixing commit. The vulnerability database is accessible through an online interface, through an API, and through downloadable database snapshots. 

\subsection{Dataset Used}

In our analysis we used the Swift library dependency network dataset compiled by Rahkema et al. \cite{rahkema-msr, rahkema2022quality} containing information on libraries from the three package managers used in iOS development. The dataset consists of nodes and relationships. In our analysis, we are interested in the following types of nodes:

\begin{itemize}
    \item App (analysed library version)
    \item Library (library version)
    \item LibraryDependency (library dependency declaration from manifest file)
    \item Vulnerability (publicly reported vulnerability)
\end{itemize}

and the following relationships:

\begin{itemize}
    \item (App) - [IS] $\rightarrow$ (Library)
    \item (App) - [CHANGED\_TO] $\rightarrow$ (App)
    \item (Library) - [LIBRARY\_DEPENDS\_ON] $\rightarrow$ (Library)
    \item (App) - [DEPENDS\_ON] $\rightarrow$ (LibraryDefinition)
    \item (Library) - [:HAS\_VULNERABILITY] $\rightarrow$ (Vulnerability)
\end{itemize}

This dataset is a node database, allowing for easy querying of node chains, which we use to find dependency chains between library versions. We give a more detailed description of the dataset in our technical report \cite{rahkema2022quality}. The dataset contains data on 60533 libraries, 572131 library versions and 23419 dependencies between libraries. 

\section{Method} \label{sec:method}
In this section, we first present the research questions that guided our study. Then we describe what analyses we conducted to answer each of the research questions. The jupyter notebook containing the scripts used in our analyses can be found on GitHub\footnote{\url{https://github.com/kristiinara/VulnerabilityPropagationAnalysis}}.

\subsection{Research Questions}

Our goal is a) to understand the scope of the library dependency network affected by vulnerabilities, b) if vulnerable dependencies could be effectively fixed via upgrading, and c) if there is enough public information available about these vulnerabilities such that the functionality of existing tools could be complemented with more detailed yet lightweight vulnerability analyses. Guided by this goal, we formulate three research questions.
\begin{itemize}
    \item RQ1: How frequently do libraries have vulnerable dependencies?
    \item RQ2: How many vulnerabilities could be fixed via upgradeing?
    \item RQ3: How precise could the vulnerability analysis be made given public information? 
\end{itemize}

In the following, we explain the underlying rationale of each of the three research questions. \\

\textbf{RQ1: Libraries with Vulnerable Dependencies} 

To better understand the risks imposed by vulnerabilities in the library dependency network, we ask how libraries in Carthage, CocoaPods and Swift PM are affected by vulnerabilities. To get started, it is necessary to investigate which libraries have publicly reported vulnerabilities. We are aware that the actual number of vulnerabilities will be higher, as not every vulnerability is publicly reported or even detected. Nevertheless, it is reasonable to look at publicly reported vulnerabilities instead of running a vulnerability scanner, to avoid the multitude of false positive results that these tools usually produce. Looking at publicly reported vulnerabilities we can be reasonably certain that these vulnerabilities are true positives and no manual double-checking is required. We expect to find publicly reported vulnerabilities in a certain amount of third-party libraries of the Swift ecosystem.

Yet, this is not sufficient. Since we expect vulnerabilities to spread through dependency chains, we analyse the library dependency network, i.e., the occurrences and lengths of dependency chains along which vulnerabilities might propagate. 

In addition, we refine our analysis by including information about the predominantly used project language of the vulnerable library and the severity level of the vulnerability. Libraries in the Swift ecosystem can be written in different languages. The most common languages are Swift, Objective-C, C and C++ \cite{dominguezusage}, with Swift and Objective-C covering the vast majority of the libraries. We expect the vulnerable libraries to have a similar distribution of languages as the rest of the ecosystem.\\ 

\textbf{RQ2: Vulnerable Dependencies Fixed via Upgrading}

The simplest way to fix a dependency to a vulnerable library version is to upgrade to a library version where the vulnerability is fixed, if such a fix exists. Given that developers are wary of upgrading their library dependencies \cite{zimmermann2019small, li2021pdgraph, zerouali2022impact} our hypothesis is that, as in other programming language ecosystems, many dependencies to vulnerable libraries remain unchanged although an easy fix is possible via upgrading the library dependency version. To check our hypothesis, we analyse how often vulnerable dependencies could have been fixed by upgrading the library dependency version. \\ 

\textbf{RQ3: Precision of Public Vulnerability Information}

Tools exist that can find dependencies to vulnerable libraries when using CocoaPods, Carthage or Swift Package Manager \cite{rahkema-msr}. There are, however, no tools for Swift and Objective-C that could perform more detailed analyses and determine if a vulnerability from a library dependency really affects the program. For such analyses it would either be necessary to have data on where the vulnerability is located in the library or an extensive analysis of the vulnerable library would be needed. Our goal is to check if information about the location of a vulnerability in the code is publicly available for the reported vulnerabilities in the Swift ecosystem. 

\subsection{Data Analysis}\label{section:dataanalysis}

In this section we describe what we do to answer the three research questions. 

\subsubsection{RQ1: Libraries with Vulnerable Dependencies}

To understand how vulnerable library versions may impact other libraries, we first find all library versions that are connected to vulnerable library versions 
through DEPENDS\_ON chains. A dependency chain of length zero implies that the library version itself is vulnerable. A dependency chain of length one implies that the library version has a direct dependency to a vulnerable library version. Dependency chains longer than one imply that the library version has a transitive dependency to a vulnerable library version. 

For each library version that depends on a vulnerable library we find the shortest path to a vulnerable library version. We do this, because we assume that the risk of using vulnerable code is higher when the dependency chain is the shortest. We then report the number of libraries for each dependency chain length by filtering out duplicate library names. The resulting numbers indicate how many libraries have publicly reported vulnerabilities and how many libraries depend on vulnerable libraries (either through direct or transitive dependencies).

Additionally, we analyse how the language of the vulnerable library and the severity level of the vulnerability is associated with how far the vulnerabilities spread in the library dependency network. We gather library dependency chains for libraries that depend on vulnerable library versions and plot the number of affected libraries for each dependency level. For libraries with multiple dependencies to vulnerable libraries, we count the library on each dependency level where it depends on a vulnerable library version. We first plot the dependency level graph distinguished by the programming language and then by the severity level of the vulnerability. 

The language of the library is determined by querying the main project language from GitHub. 

\subsubsection{RQ2: Vulnerable Dependencies Fixed via Upgrading}

Figure \ref{fig:dependency-chains} in Section \ref{sec:background} showed five dependency chains of which three corresponded to vulnerable dependencies: ABC1, AB1 and ABC2. For dependency chain ABC2, Library C could have resolved the vulnerable dependency by upgrading the dependency to Library B from version 1 to version 2. For RQ2, we will study how often such chains to vulnerable dependencies could have been fixed via upgrading the dependency version. 

For this analysis we first filter out library dependencies where the package manager resolution file was missing. These dependency versions were calculated based on the manifest file and are therefore not suitable for upgradeability analysis. 

To analyse how vulnerable dependencies could be fixed via upgrading, we first identify all dependency chains to vulnerable library versions. For each of these chains we then check if a newer version of the direct dependency (like B:v2 for dependency chain ABC2 in Figure \ref{fig:dependency-chains}) exists that is not dependent on the vulnerable library version. The process of finding the newer version of the direct dependency takes into account release times for each of the library versions such that the release time of the dependency has to always be before the release time of the dependent. This means that in Figure \ref{fig:dependency-chains} it would have been possible to upgrade the dependency chain ABC2, but not the dependency chain ABC1 because B:v2 was released after C:v1. 

For each dependency level we plot the number of dependency chains that could have been fixed via an upgrade and the number that could not have been fixed via an upgrade. Additionally we count how many library dependencies could have been fixed for each vulnerability severity level and vulnerable library programming language. 

The above analysis shows the upgradeability of vulnerable dependencies over the whole time frame of the dataset. To understand the potential impact of upgrades to the most recent state of the library dependency network, we also analyse for how many of the latest versions of libraries vulnerable dependencies could have been fixed via upgrading. 

\subsubsection{RQ3: Precision of Public Vulnerability Information}

For each vulnerability, we check the public vulnerability description on NVD and record if it contains information about the class or method that contains the vulnerability. Additionally we check, if available, the patch link to see if the patch of the vulnerability reveals where the vulnerability was fixed in the code. 

\section{Results}\label{sec:results}
In the following, we present the answers to our research questions one by one.

\subsection{RQ1: Libraries with Vulnerable Dependencies}

We found a total of 149 vulnerabilities in 61222 libraries. This corresponds to 24.3 vulnerabilities per 10000 libraries. We found that only 5.9\% of connected libraries had dependencies to vulnerable library versions. For 3\% of connected libraries, even the latest version of the library had a dependency to a vulnerable library version. 

In the following, we present in more detail our results on how vulnerabilities propagate through the Swift library dependency network. Furthermore, we show how library project language and vulnerability severity are associated with vulnerability propagation. 

\begin{figure}[htbp]
\centerline{\includegraphics[width=1.0\columnwidth]{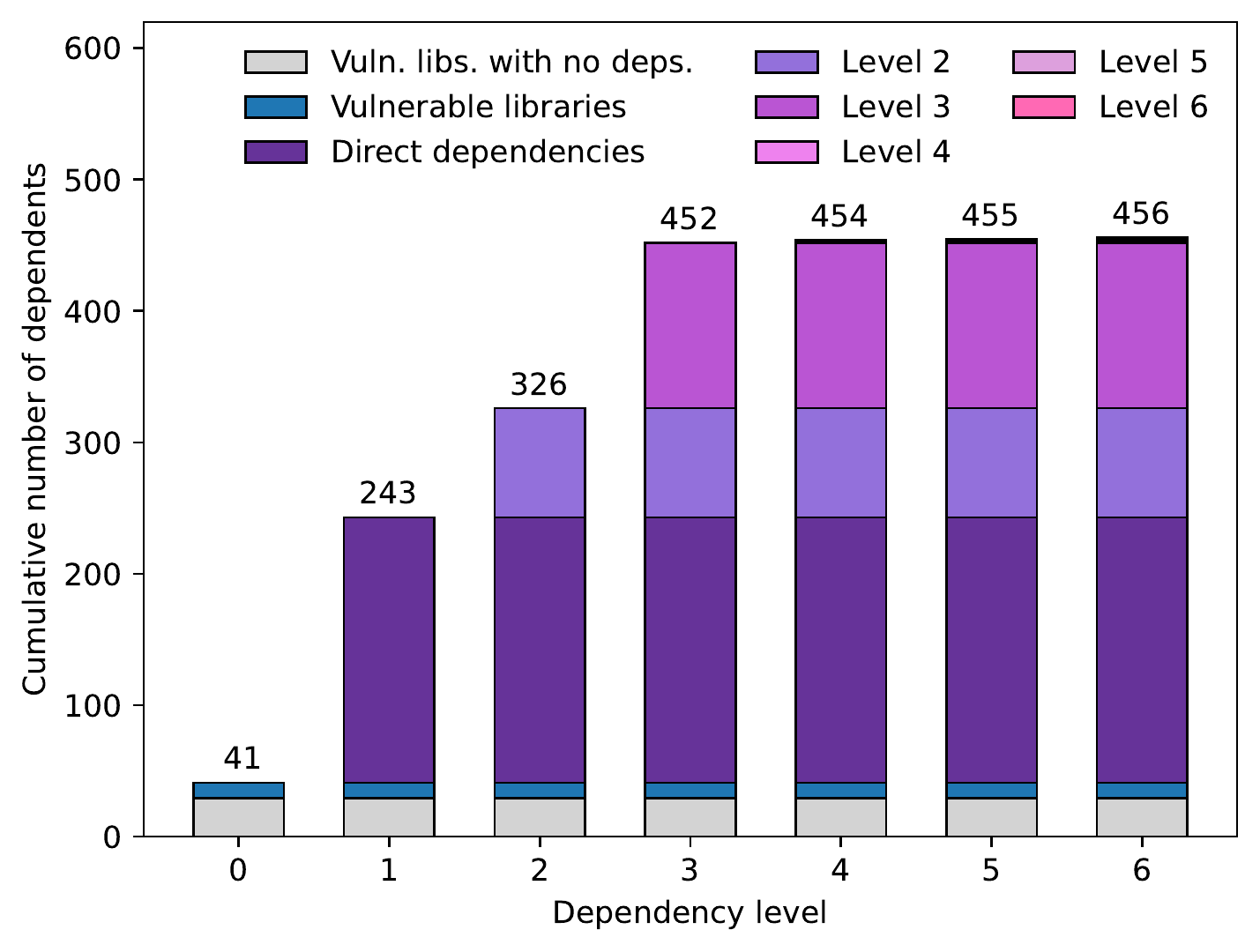}}
\caption{
Cumulative number of libraries affected by vulnerabilities for each dependency level classified by the shortest dependency level to a vulnerable library version for each library. 
}
\label{fig:vulnerable-dependencies}
\end{figure}

Figure \ref{fig:vulnerable-dependencies} shows how publicly reported vulnerabilities propagate through the Swift library dependency network. There are 41 libraries with publicly reported vulnerabilities (dependency tree level 0). Of those libraries only 12 have dependents. There are 202 libraries without a publicly reported vulnerability that have a direct dependency to at least one vulnerable library version (dependency tree level 1). A considerable number of libraries are added on level two (83) and level three (126). Libraries with dependencies to multiple vulnerable library versions are counted at the lowest dependency tree level where a dependency to a vulnerable library version exists. In total, 415 libraries have dependencies to vulnerable library versions, and 456 libraries are affected by publicly reported vulnerabilities in total, if we include libraries that are vulnerable themselves. 
Moreover, we can say that in case a library has at all a (possibly indirect) dependency to a vulnerable library version, then the longest chain to the first vulnerable library version has at most six levels in the dependency tree.

Table \ref{tab1e:language} shows the results of the analysis that explored whether the programming language in which a library is written has an influence on how vulnerabilities spread through the dependency network. We determined the programming language of each vulnerable library based on data from GitHub on the main project language of the library. Table \ref{tab1e:language} indicates that most vulnerabilities originate in libraries written in C (88) and C++ (24). Libraries written in Swift and Objective-C contribute only 19 and three vulnerabilities, respectively. 

\begin{figure}[htbp]
\centerline{\includegraphics[width=1.0\columnwidth]{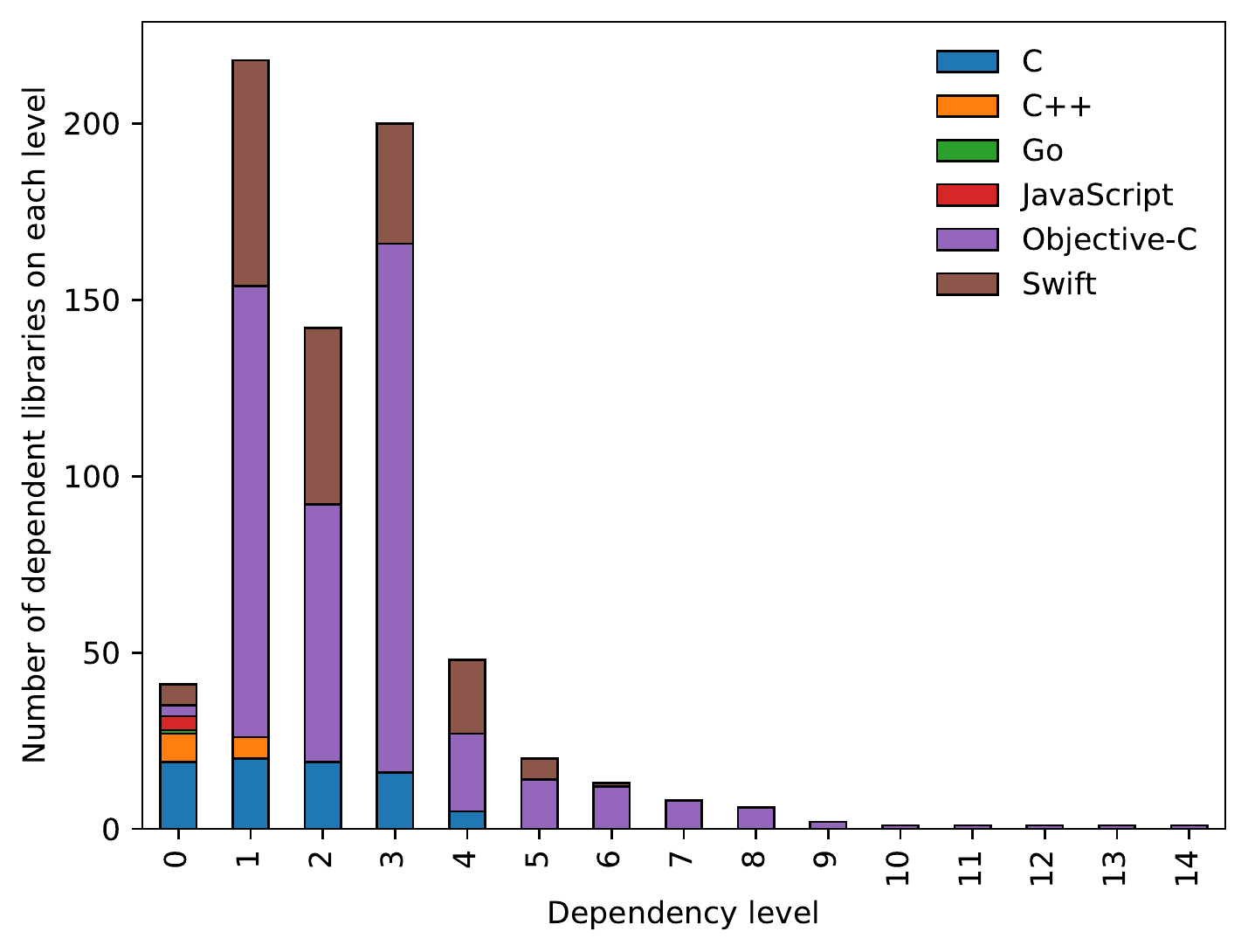}}
\caption{Number of libraries affected by vulnerabilities for each dependency level classified by main project language.}
\label{fig:vulnerable-dependencies-lang}
\end{figure}

However, the highest impact on the Swift ecosystem comes from vulnerabilities in libraries written in Swift and Objective-C. Table \ref{tab1e:language} shows that vulnerable libraries written in Swift and Objective-C have significantly more dependents (98 for Swift and 313 for Objective-C) than projects written in other programming languages (56 and 14 for C and C++). Figure \ref{fig:vulnerable-dependencies-lang} shows how far vulnerabilities from libraries written in the different languages spread across the dependency network. 
Although there are some libraries that have dependencies to libraries written in C and C++, there are significantly longer dependency chains to libraries written in Swift and, especially, to libraries written in Objective-C. In the case of Objective-C dependency chains to vulnerable library versions can have up to 14 levels of indirection. Differently to Figure \ref{fig:vulnerable-dependencies}, libraries in Figure \ref{fig:vulnerable-dependencies-lang} are counted on each level of indirection they occur. 

\begin{table}[htbp]
\caption{Vulnerabilities by project language}
\begin{center}
\begin{tabular}{l c c c}
\toprule
Project language & vulnerabilities & libraries& dependent libraries\\
\midrule
C & 88 &  19 & 56 \\
C++ & 24 & 8 & 14 \\
\textbf{Swift} & \textbf{19} & \textbf{6} & \textbf{98} \\
Go & 12 & 1 & 1 \\
JavaScript & 4 & 4 & 4 \\
\textbf{Objective-C} & \textbf{3} &\textbf{3} & \textbf{313} \\
\toprule
\end{tabular}
\label{tab1e:language}
\end{center}
\end{table}

In the following, we present our results on how vulnerabilities of different severity propagate throughout the library dependency network. Vulnerabilities have four levels of severity: CRITICAL, HIGH, MEDIUM and LOW. Table \ref{tab1e:severity} provides information on the distribution of severity levels of the vulnerabilities found in the Swift ecosystem, as well as dependent libraries affected by these vulnerabilities. Most vulnerabilities (80) are of level HIGH, 31 and 37 vulnerabilities are CRITICAL and MEDIUM, respectively. Only one vulnerability has the level LOW. Most libraries (353) are affected by MEDIUM level vulnerabilities through dependencies. 

\begin{figure}[htbp]
\centerline{\includegraphics[width=1.0\columnwidth]{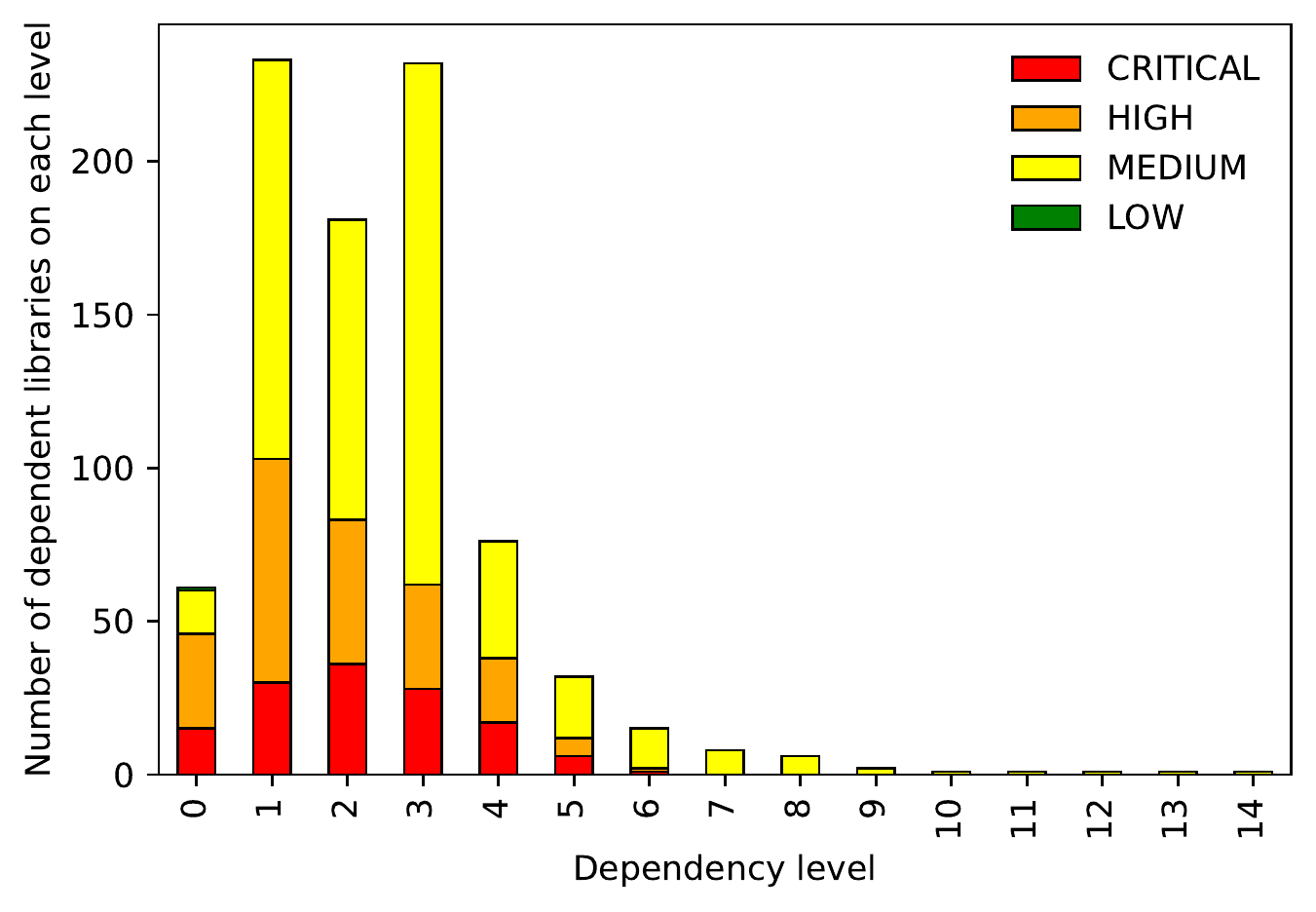}}
\caption{Number of libraries affected by vulnerabilities for each dependency level classified by severity level of the vulnerability.}
\label{fig:vulnerable-dependencies-severity}
\end{figure}

Figure \ref{fig:vulnerable-dependencies-severity} shows how vulnerabilities with different severity levels propagate through the library dependency network. Vulnerabilities with severity level MEDIUM propagate the furthest through the dependency network. However, vulnerabilities with severity level CRITICAL and HIGH can both be observed with levels of indirection in the dependency tree up to Level 5. 

\begin{table}[htbp]
\caption{Vulnerabilities by severity}
\begin{center}
\begin{tabular}{ l c c c}
\toprule
Vulnerability severity & vulnerabilities & libraries& dependent libraries\\
\midrule
CRITICAL & 31 &  15 & 73 \\
HIGH & 80 & 31 & 136 \\
MEDIUM & 37 & 14 & 353 \\
LOW & 1 & 1 & 1 \\
\toprule
\end{tabular}
\label{tab1e:severity}
\end{center}
\end{table}

\subsection{RQ2: Vulnerable Dependencies Fixed via Upgrading}

To answer RQ2, we analyse how many vulnerable dependencies could have been fixed via a dependency upgrade at the time a library version was released. For the upgradability analysis we require that the version data for direct dependencies originates from package manager resolution files and that the direct dependency is to a library included in the set of libraries available for our analysis. After filtering out dependencies that did not meet our criteria 341 out of 415 libraries with vulnerable dependencies remain. 

\begin{figure}[htbp]
\centerline{\includegraphics[width=1.0\columnwidth]{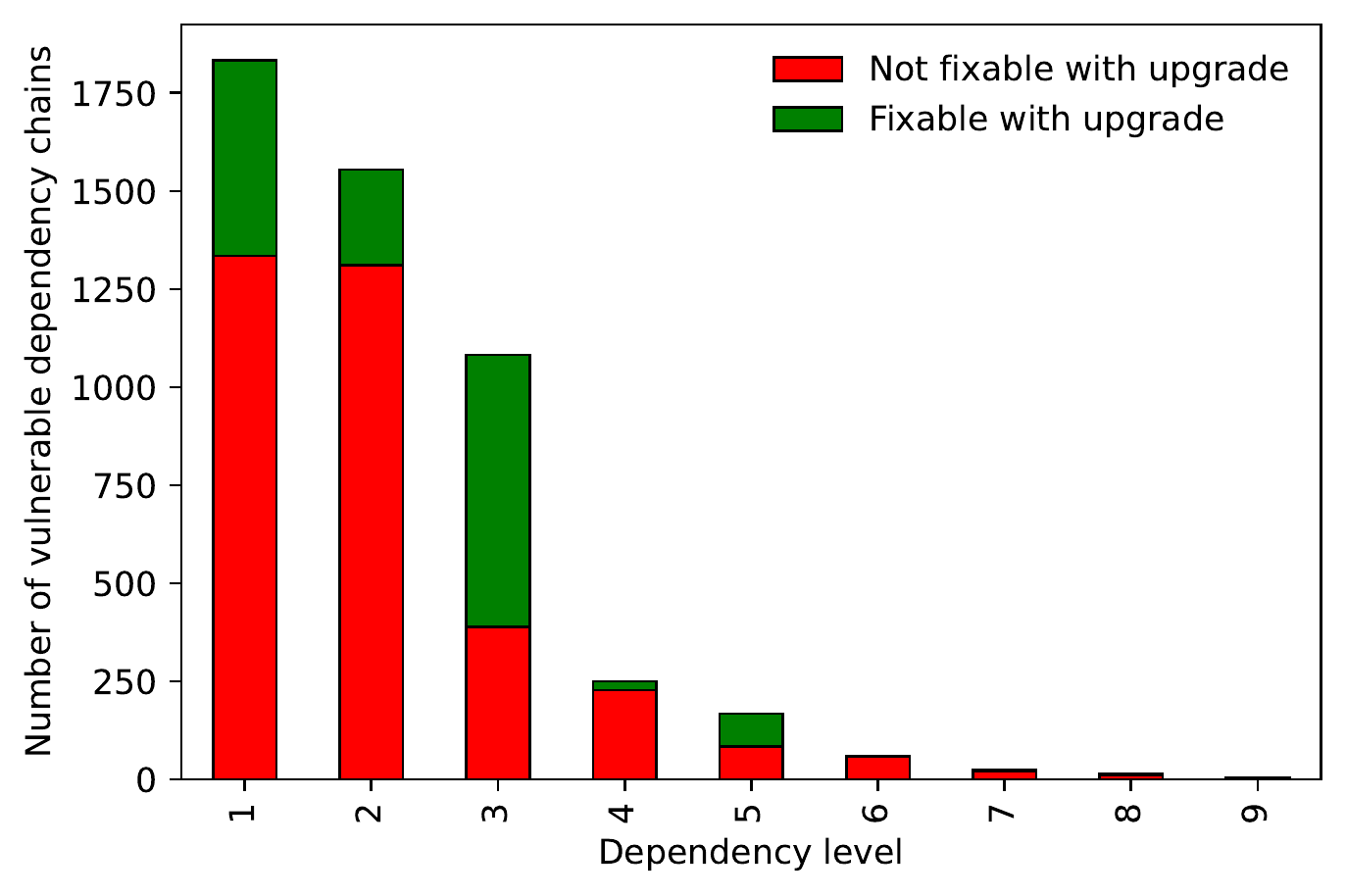}}
\caption{Number of dependency chains to vulnerable library versions that could be fixed (green) and not fixed (red) by an upgrade of the first dependency in the dependency chain. The numbers are shown for each dependency level. 
}
\label{fig:vulnerable-dependencies-upgrade}
\end{figure}

First, we analyse how updating direct dependencies would have fixed a vulnerable dependency for different levels of indirection. Figure \ref{fig:vulnerable-dependencies-upgrade} shows that 27\% (498 of 1833 in total) of vulnerable direct dependencies could have been fixed via an upgrade. Furthermore upgrading direct dependencies would also have fixed 16\% (244 of 1555 in total) of second level vulnerable dependencies and 64\% (694 of 1082 in total) of third level vulnerable dependencies. Note that the levels of dependency tree in Figure \ref{fig:vulnerable-dependencies-upgrade} are greater than 0 and less than 10. They must be greater than 0 because there are no dependencies at level 0. There is no data beyond level 9 as the data on those library chains happened to not be compatible with the upgradability analysis. 

Next, we analyse how many vulnerable dependencies could have been fixed via upgrading depending on the severity of the vulnerability. Table \ref{tab1e:severity-fixes} shows that over all dependency chains the probability of fixing the vulnerability via a dependency upgrade is around 30\%. However, if we look at the latest version of each library, the percentage of fixing dependencies to critical vulnerabilities via upgrading is 71\%, fixing dependencies to vulnerabilities of level HIGH is 52\% and fixing dependencies to vulnerabilities of level MEDIUM is 39\%.  

\begin{table}[htbp]
\caption{Vulnerable dependency fixes by severity}
\begin{center}
\begin{tabular}{l|cc|cc}
\toprule
Vulnerability & \multicolumn{2}{c|}{all versions}&\multicolumn{2}{c}{latest version} \\
 severity & fixed & not fixed & fixed & not fixed\\
\midrule
CRITICAL & 31\% &  69\%  &71\% & 29\%\\
HIGH & 33\% & 67\% &52\% & 48\%\\
MEDIUM & 30\% & 70\% &39\% & 61\%\\
\toprule
\end{tabular}
\label{tab1e:severity-fixes}
\end{center}
\end{table}

Finally, we explore whether there are differences in the percentages of fixing vulnerabilities via upgrading between the project languages of the vulnerable libraries. Table \ref{tab1e:language-fixes} shows percentages of vulnerable dependencies being fixed via upgrading for each of the four most prominent languages. Over all library versions, the probability of fixing a vulnerable dependency is around 30\%, with the exception of C++ where the probability is considerably smaller. For the latest versions of each library the probability of fixing a vulnerable dependency via an upgrade is highest for C (67\%) and Swift (60\%), and lowest for Objective-C (38\%) and C++ (33\%). 

\begin{table}[htbp]
\caption{Vulnerable dependency fixes by project language}
\begin{center}
\begin{tabular}{l|cc|cc}
\toprule
Project & \multicolumn{2}{c|}{over all versions}&\multicolumn{2}{c}{latest version only} \\
 
language & fixed & not fixed & fixed & not fixed\\
\midrule
C & 24\% &  76\% & 67\% & 33\%\\
C++ & 6\% & 94\%& 33\% & 67\%\\
Objective-C & 30\% & 70\% & 38\%& 62\%\\
Swift & 36\% & 64\% & 60\% & 40\%\\
\toprule
\end{tabular}
\label{tab1e:language-fixes}
\end{center}
\end{table}

Looking at the success rates of fixing a vulnerable dependency via upgrading from the point of view of the different vulnerabilities we see that for 25\% of vulnerabilities the success of upgrading is over 89\% and for another 25\% of vulnerabilities the failure of fixing the dependency via an upgrade is over 94\%. These numbers indicate that fixing a vulnerable dependency via upgrading is very successful for some of the vulnerabilities and not possible for others.

\subsection{RQ3: Precision of Public Vulnerability Information}

Our answer to RQ3 is based on checking whether the public descriptions of vulnerabilities in the Swift ecosystem include information on the class or method that contains the described vulnerability. This kind of detailed information could be used to fine tune the analysis and detection of vulnerable dependencies by identifying the piece of code that contains the vulnerability. In situations where a library is dependent on a vulnerable library version it might be good to know whether the code of the vulnerable library is used by the dependent library. Analysing the description of each vulnerability and including data from patch links showed that most vulnerability descriptions do not include detailed enough information to determine the vulnerable class or method. Table \ref{tab1e:precision} shows that not a single vulnerability in a project written in Swift specified both the vulnerable method and the vulnerable class. Similarly, very little information is available about vulnerabilities in projects written in Objecive-C. There is more information on vulnerabilities in projects written in C and C++ but these vulnerabilities also affect significantly less libraries in the Swift ecosystem. 

\begin{table}[htbp]
\caption{Precision of public information on vulnerabilities}
\begin{center}
\begin{tabular}{lcccc}
\toprule
Project language & vulnerabilities & method & class & both\\
\midrule
Swift & 19 & 1 & 7 & 0 \\
Objective-C & 3 & 1 & 1 & 1 \\
C & 88 & 43 & 29 & 16 \\
C++& 24 & 18 & 15 & 12 \\
\toprule
\end{tabular}
\label{tab1e:precision}
\end{center}
\end{table}

\section{Discussion}\label{sec:discussion}

\subsection{RQ1: Libraries with Vulnerable Dependencies}

The total amount of 149 vulnerabilities in 61222 libraries in the Swift ecosystem corresponds to 24.3 vulnerabilities per 10000 libraries. This ratio is much higher than that for npm where Zimmermann et al. found the ratio to be around 8 in 2018 \cite{zimmermann2019small}. The difference between the two ecosystems could be due to the high number of very small libraries in npm. In contrast, Li et al. found the ratio to be 113.5 for Java projects \cite{li2021pdgraph}. The Java ecosystem is older and might have larger libraries, but we do not have a definite reason for the big difference. 

Our results show that only 5.9\% of connected libraries have dependencies to vulnerable library versions. For 3\% of connected libraries, its latest version is still dependent on a vulnerable library version. In contrast, Düsing et al\cite{dusing2021analyzing} found that 9\% of libraries in npm had direct dependencies to vulnerable libraries. Zimmermann et al. \cite{zimmermann2019small} found that 40\% of npm projects they studied depended (directly or transitively) on vulnerable libraries. Alfadel et al. \cite{alfadel2020threat} found that 67\% of all npm applications had at least one vulnerable direct dependency. Zerouali et al. \cite{zerouali2022impact} found that for more than 15\% of npm and RubyGems libraries the latest version of the library is directly dependent on a vulnerable library version. Additionally, they found that for 42.1\% of all npm libraries and for 30\% of RubyGems libraries the latest version of the library had a transitive dependency on a vulnerable library version. Therefore, in comparison to other ecosystems, the Swift ecosystem is considerably less affected by vulnerable library dependencies. A possible reason could be that libraries in the Swift ecosystem have less dependencies on average than libraries in other ecosystems, such as npm. Another possibility is that there are less vulnerabilities reported for the libraries in the Swift ecosystem but our analysis shows that this is not true, at least in comparison to npm. 

Looking at the severity of the vulnerabilities, the vulnerabilities with severity level MEDIUM spreads the most in the library dependency network. A possible explanation is that vulnerabilities with a MEDIUM severity level are not taken as seriously as vulnerabilities with higher severity levels and, therefore, are able to exist longer and spread further in the library dependency network. 

Most vulnerabilities in the Swift ecosystem originate form libraries written in C and C++. When looking at the impact on the whole library dependency network, however, vulnerabilities in libraries written in Swift and Objective-C spread considerably farther. Libraries written in Swift and Objective-C have more dependents and therefore a higher impact on the overall library dependency network. Dom{\i}nguez-Alvarez et al. \cite{dominguezusage} found that most libraries available through the CocoaPods package manager are written in Swift and Objective-C. It might be, that libraries written in C and C++ are very specialized and therefore not used by many other libraries. 

\subsection{RQ2: Vulnerable Dependencies Fixed via Upgrading}

Overall, around 30\% of vulnerable dependencies could have been fixed via an update of direct dependencies. Surprisingly, there is not much difference between vulnerability severity levels when looking at upgradability over all library versions. When looking at the latest version of each library, however, our results show that vulnerabilities with severity level CRITICAL could have been fixed in 70\% of the cases. This is a strong indication for developers to keep up with library dependency upgrades as a means to avoid dependence on vulnerable libraries. If upgrading to each new version is not possible, developers should at least check if their dependencies have publicly reported vulnerabilities, for example using automated tooling such as \cite{rahkema2022checker}.

\subsection{RQ3: Precision of Public Vulnerability Information}

Currently, tools exist that can be used to check for vulnerable dependencies when using CocoaPods, Carthage or Swift Package Manager\cite{rahkema2022checker}. There are, however, no tools for the Swift ecosystem that could check if a vulnerability in a dependency really affects the developed application. Existing tools could be extended if detailed information about the exact code location of a vulnerability was available. Our results suggest that NVD does not include enough detailed information on vulnerabilities in libraries written in Swift and Objective-C. Therefore, the best solution for developers is to upgrade to a version of the library where the vulnerability has been fixed - if such a version is available.

\section{Threats to Validity}\label{sec:threats}

In this section we discuss the potential limitations of our analyses, focusing on construct, internal, and external validity threats. 

\subsection{Construct Validity}
In our analysis, we assume that every vulnerable dependency implies that the dependent library is (indirectly) vulnerable as well. However, the presence of vulnerable dependencies does not necessarily imply that the library is actually vulnerable. In a preliminary study \cite{zapata2018towards} Zapata et al. analysed dependencies of 60 projects using the npm package manager and showed that most projects with vulnerable dependencies do not actually use the vulnerable code. Hejderup et al. \cite{hejderup2021pr} analysed libraries written in Rust and showed that not all resolved dependencies are really called, which means that dependencies to vulnerable libraries might not necessarily affect the library itself. Given that our results show that relatively few libraries depend on vulnerable libraries in the library dependency networks of the Swift ecosystem, a more detailed analysis would not affect this conclusion. A detailed analysis of call graphs might reduce the percentage of libraries with dependencies to vulnerable libraries even further. However, as we show in our answer to RQ3, the data needed for such an analyis is often not available. 

Our analyses using information about the programming languages in which the vulnerable libraries are written depends on the information provided by GitHub about the main programming language. A library could, however, be written in several programming languages and the vulnerability itself could be located in code that was written in a programming language different from the main programming language. To understand the level of correctness of the information provided by GitHub in the context of our analysis, for those vulnerabilities where class/file level data was available, we also checked the language of the class/file and compared it to the main programming language indicated by GitHub. In only two of the 92 cases where such information is available, the vulnerability is located in a file written in a different language as the main programming language of the library. None of these two cases occurs in libraries classified as written in Objective-C or Swift. 

\subsection{Internal Validity}

The analysed dependency data relies on package manager resolution files. Not every library that uses a package manager includes the corresponding resolution files in the repository. For such repositories the package manager manifest files are used to resolve the dependencies following Rahkema et al. \cite{rahkema-msr}. 

Building the dependency graph by only declaring the exact version of a dependency means that transitive dependencies could in practice be resolved differently. When a transitive dependency is resolved at a later date then it is possible that the actual version of the transitive dependency would not match the version in the dataset. The data on the version ranges does, however, exist in the dataset and could be checked as future work. 

For the vulnerability data we rely on data from NVD. This means that we need to trust that the data is correct. It is possible that there are incorrect entries if they have not been checked by third parties. We do, however, believe the data to be reliable as it is an official and public database which is continuously reviewed and maintained. 

\subsection{External Validity}

We claim that our results hold for all open source libraries in the library dependency networks of the Swift ecosystem, i.e. all open source libraries that are available through CocoaPods, Carthage and Swift PM. CocoaPods, Carthage and Swift PM are the only package managers used in the Swift ecosystem. 

The dataset we analyzed represents the Swift ecosystem in the period from September 2011 to December 2021. We make no claims regarding how the vulnerability propagation might evolve in the future.

The vulnerability data in the dataset is based on public data from the NVD. When using other vulnerability databases, for example Snyk, the results might be different. Vulnerabilities in NVD are publicly reported, which adds to the trustworthiness of the data. 

\section{Conclusion}\label{sec:conclusion}

We studied the vulnerability propagation in the library dependency network controlled through the package managers used in iOS development, and analysed how developers could protect against vulnerabilities from third party libraries. 

We found that the Swift ecosystem is affected less by public vulnerabilities than other ecosystems such as npm. The spread of vulnerabilities is smaller, probably due to less dependencies on average. We showed that upgrading direct dependencies is an effective way to fix vulnerable dependencies and, as long as no tools exist that correctly analyse whether the vulnerable code is actually called, the best option for developers is to regularly upgrade library dependency versions. Currently, no tools exist for Swift and Objective-C that are able to perform such a detailed analysis, and we showed that there is not enough public data available to implement such tools easily. 

\bibliographystyle{IEEEtran}
\bibliography{bibliography}

\end{document}